\documentclass[aps,showpacs,floatfix,floats,prb,amsmath,amssymb,twocolumn,reprint]{revtex4}
\usepackage{graphicx}
\usepackage{color}
\usepackage{ulem}
\begin{document}

\title{ Field-induced decay dynamics in square-lattice antiferromagnet}

\author{
M. Mourigal,$^{1,2}$ M. E. Zhitomirsky,$^{3,4}$  and
A. L. Chernyshev,$^{5,4}$
}
\affiliation{
$^1$Institut Laue-Langevin, BP 156, 38042 Grenoble Cedex 9, France  \\
$^2$Laboratory for Quantum Magnetism, \'Ecole Polytechnique F\'ed\'erale
de Lausanne (EPFL), 1015 Lausanne, Switzerland  \\
$^3$Service de Physique Statistique, Magn\'etisme et
Supraconductivit\'e, UMR-E9001 CEA-INAC/UJF, 17 rue des Martyrs,
38054 Grenoble Cedex 9, France \\
$^4$Max-Planck-Institut f\"ur Physik Komplexer Systeme, N\"othnitzer str.\
38,  D-01187 Dresden, Germany \\
$^5$Department of Physics and Astronomy, University of California, Irvine,
California 92697, USA
}
\date{September 3, 2010}

\begin{abstract}
Dynamical properties of the square-lattice Heisenberg antiferromagnet in
applied magnetic  field are studied for arbitrary value $S$ of the spin. 
Above the threshold field for two-particle decays, the standard spin-wave 
theory yields singular corrections to the excitation spectrum
with logarithmic divergences for certain momenta.
We develop a self-consistent approximation applicable for $S \geq 1$, which
avoids such singularities and provides regularized  magnon decay rates.
Results for the dynamical structure factor obtained in this approach 
are presented for $S = 1$ and  $S = 5/2$.
\end{abstract}
\maketitle

\section{Introduction}

The square-lattice Heisenberg antiferromagnet is an important model system
in quantum magnetism.  \cite{Manousakis91} Theoretical studies of this model
have provided deep insights into the role of low-dimensionality in the static
and dynamical properties of the many-body spin systems. Recently, there has been
a growing interest in the effect of magnetic field in the dynamics of quantum
antiferromagnets. The strong-field regime is now reachable for
a number of layered square-lattice materials with moderate strength of exchange
coupling between spins. \cite{Woodward02,Lancaster07,Coomer07,Tsyrulin09}
In addition, new field-induced dynamical effects can be present in the
antiferromagnets with other lattice geometries \cite{Coldea02,Nakatsuji05}
as well as in the gapped quantum spin systems near the condensation field for
triplet excitations.
\cite{Affleck91,Affleck92,Mila98,Giamarchi99,Nikuni00,Jaime04,Regnault06}

\begin{figure}[t]
\centering
\includegraphics[width=0.75\columnwidth]{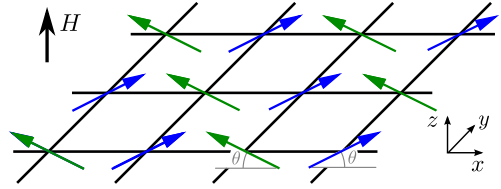}
\caption{(Color online) Canted spin structure of
the square-lattice Heisenberg antiferromagnet in
external magnetic field.
}
\label{canted_structure}
\end{figure}

The ground-state properties of the square-lattice antiferromagnet in a finite
field conform with the semiclassical picture of spins gradually tilted towards
the field  direction, \cite{Zhitomirsky98}  see Fig.~\ref{canted_structure}.
On the other hand, excitation spectrum and dynamical properties are expected
to undergo rather dramatic transformation. \cite{Zhitomirsky99,Olav08,Luscher09}
Following the prediction of
the field-induced spontaneous magnon decays in Ref.~\onlinecite{Zhitomirsky99},
there is an ongoing search for suitable spin-1/2 square-lattice materials
\cite{Woodward02,Lancaster07,Coomer07,Tsyrulin09} to investigate such an effect.
The existence of substantial damping in the magnon spectrum of the square-lattice
Heisenberg model in a field has been recently verified by the Quantum  Monte
Carlo \cite{Olav08} (QMC) and the Exact Diagonalization \cite{Luscher09} numerical study.
Other theoretical aspects of the behavior of the quantum square-lattice
antiferromagnet (SAFM) in applied field have also been addressed.
\cite{Kreisel08,Syromyat09,Chernyshev09b}

In this paper, we extend previous work of two of us \cite{Zhitomirsky99}
and provide a comprehensive theoretical investigation of the dynamics of
the nearest-neighbor Heisenberg SAFM including detailed calculation of
the $1/S$ correction to the energy spectrum in external field,
kinematic analysis of the field-induced two-magnon decays, and self-consistent
treatment of magnon decay rates for systems with  $S\geq 1$.

The model spin Hamiltonian is
\begin{equation}
\hat{\cal H}  =  J \sum_{\langle ij \rangle} \mathbf{S}_i\cdot\mathbf{S}_j
           -   H \sum_{i}  S_i^{z}  \ ,
\label{2DHAFSL}
\end{equation}
where $J$ stands for the nearest-neighbor exchange coupling constant
and $ H$ is the external magnetic field in units of $g\mu_B$.

The standard spin-wave theory works quite well for the SAFM in zero field.
\cite{Manousakis91,Hamer92,Igarashi92,Canali93,Igarashi05}
Surprisingly enough, in high magnetic fields one encounters
strong singularities in the spin-wave corrections to the dynamical
properties of the SAFM, which arise due to spontaneous magnon decays
above the threshold field $H^*\approx 0.75H_s$. \cite{Zhitomirsky99}
Singular behavior of the spectrum signifies a breakdown of the perturbative
$1/S$ expansion and requires a regularization. The aim of this work
is to develop a self-consistent approximation which yields the
spectrum that is free from the essential singularities.

Within the conventional spin-wave approach, the gapless
Goldstone branch is preserved only if all quantum corrections to the spectrum
of the same order in $1/S$ are taken into account.
This represents an obvious difficulty for any self-consistent calculation, which
typically involves summation of a certain infinite
subset of perturbation series. The basic idea of the present work is to neglect
the real part of the spectrum corrections completely and to perform
the self-consistent calculation only in the imaginary part of the magnon self-energy.
Such an approach is justified if the real part of corrections is small, which is the case for $S\geq 1$ for the model (\ref{2DHAFSL}). Utilizing this
self-consistent scheme we obtain explicit results for the magnon decay rates
and the dynamical structure factor of SAFMs with $S=1$ and $S=5/2$.

The paper is organized as follows. Technical details of the spin-wave
theory in magnetic field are provided in Sec.~II together with explicit
calculation of the quantum correction for the spin-1/2 SAFM.
The kinematic analysis of the field-induced  magnon decays and associated
singularities is presented in Sec.~III. The self-consistent theory
is described  in Sec.~IV, where the results for the dynamical structure factor
are also included. Sec.~V contains our conclusions. The asymptotic form for
the decay rate of low-energy magnons is derived in Appendix A and
an extension of the kinematic analysis to anisotropic antiferromagnets
is given in Appendix B.

\section{Spin-Wave Expansion}
\label{formalism}

In this section we provide necessary details of the standard spin-wave formalism
as applied to two-sublattice  antiferromagnets in external magnetic field
at zero temperature. \cite{Zhitomirsky98,Zhitomirsky99,Chernyshev09b} The first
essential step is to quantize spin components in the rotating frame $(x_i,y_i,z_i)$
such that the local $\hat{\bf z}_i$ axis points in the direction of each magnetic
sublattice. In the case of the square-lattice model the ground state in zero
magnetic field is the collinear N\'eel order characterized by the wave-vector
$\mathbf{Q}=(\pi,\pi)$. In a finite magnetic field spins cant towards the field
direction  by an angle $\theta$, see Fig.~\ref{canted_structure}. Spin components
in the laboratory frame $(x_0,y_0,z_0)$ are related to those in the local frame by
\begin{eqnarray}
 S_i^{x_0} & = & S_i^x \sin\theta + S_i^z e^{i {\bf Q}\cdot {\bf r}_i} \cos\theta \ ,
        \quad S_i^{y_0}=S_i^{y} \ , \nonumber \\
 S_i^{z_0} & = & -S_i^x e^{i{\bf Q}\cdot{\bf r}_i} \cos\theta + S_i^z \sin\theta \ .
\label{eq:frame}	
\end{eqnarray}
The above representation allows us to introduce only
one type of bosons via the standard Holstein-Primakoff transformation. \cite{HP}
Substituting (\ref{eq:frame}) into Eq.~(\ref{2DHAFSL}) and expanding
square-roots one  obtains bosonic Hamiltonian as a sum
$\hat{\cal H}= \sum_{n=0}^\infty \hat{\cal H}_{n}$, each term being of
the $n$-th order in bosonic operators and carrying an explicit factor
$S^{2-n/2}$. For large spins this form provides the basis
for the $1/S$ expansion.

Minimization of  the classical energy $\hat{\cal H}_0$ determines
the canting angle
\begin{equation}
\sin \theta = H / H_s \quad  \textrm{with}\quad  H_s = 8JS \ .
\end{equation}
The linear terms given by $\hat{\cal H}_{1}$ vanish
for this choice of  $\theta$ and expansion begins with
the  quadratic terms
\begin{eqnarray}
\hat{\cal H}_2 & = & H\sin\theta\sum_i a_i^\dagger a_i +
JS \sum_{\langle ij \rangle}
\Bigl[\cos 2\theta\,(a^\dagger_i a_i + a^\dagger_j a_j)
\nonumber \\
& +&
\sin^2\!\theta\,( a^\dagger_i a_j  + a^\dagger_j a_i) -
\cos^2\!\theta\,( a^\dagger_i a^\dagger_j  + a_j a_i)\Bigr]\,.
\label{H2}
\end{eqnarray}
Performing successive Fourier and Bogolyubov transformations,
the latter being defined as
$$
a_{\bf k} = u_{\bf k} b_{\bf k} + v_{\bf k} b^\dagger_{-\bf k} \ ,
$$
we obtain the diagonal form of
$\hat{\cal H}_2$:
\begin{equation}
\hat{\cal H}_2 =\sum_{\bf k} \epsilon_{\bf k} b_{\bf k }^{\dagger} b_{\bf k}
 + \frac{1}{2} \sum_{\bf k} \bigl(\epsilon_{\bf k}-A_{\bf k}\bigr)
 \label{spectrum}
\end{equation}
with
\begin{eqnarray}
\label{epsilon}
&& \epsilon_{\bf k} = 4JS\sqrt{(1+\gamma_{\bf k})(1-\cos 2 \theta \gamma_{\bf k})}
= 4JS \omega_{\bf k} \  ,
\\
&& A_{\bf k} = 4 J S (1+\sin^2\!\theta\,\gamma_{\bf k})\ , \quad
\quad u_{\bf k}^2,v_{\bf k}^2 =
\frac{A_{\bf k} \pm \epsilon_{\bf k}}{2 \epsilon_{\bf k}} \ ,
\nonumber
\end{eqnarray}
and $\gamma_{\bf k} =\frac{1}{2}(\cos k_x + \cos k_y)$.
The magnon energy in the harmonic (or linear spin-wave)
approximation is given by $\epsilon_{\bf k}$.
The field-induced planar anisotropy opens a gap
at ${\bf k} = 0$, while an acoustic branch is preserved
for  $\bf k \rightarrow Q$.

\begin{figure}[t]
\centering
\includegraphics[width=0.85\columnwidth]{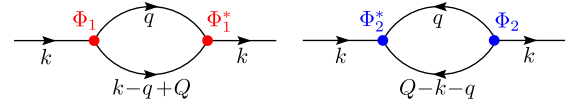}
\caption{(Color online) The lowest-order normal self-energies generated
by the decay (left) and by the source  (right) vertex.}
\label{diagrams}
\end{figure}

The  lowest-order quantum correction to the magnon energy
is determined by two types of anharmonicities, cubic and quartic.
The cubic term, which originates from
the coupling between transverse and longitudinal fluctuations in
the canted spin structure, is given by
\begin{eqnarray}
\hat{\cal H}_3 & = & \sqrt{\frac{S}{2}} \Bigl[
J\sin 2\theta
\sum_{i,j} e^{i{\bf Q}{\bf r}_i} (a^\dagger_i + a_i) a^\dagger_j a_j
 \label{H3} \\
& + &\cos\theta
\Bigl(2J\sin\theta -\frac{H}{4S}\Bigr)
\sum_{i} e^{i{\bf Q}{\bf r}_i}
\bigl(a^\dagger_i n_i + n_i a_i\bigr) \Bigr]
\nonumber
\end{eqnarray}

with $n_i = a^\dagger_i a_i$. The quartic term is
\begin{eqnarray}
\hat{\cal H}_4 & = & J\sum_{\langle ij\rangle}
\Bigl\{\frac{1}{4}\cos^2\!\theta\,\bigl[
(n_i+n_j)a_i a_j + {\rm h.\,c.} \bigr] \label{H4}  \\
 &-& n_i n_j\cos 2\theta - \frac{1}{4}\sin^2\!\theta\, \bigl[
a^\dagger_i(n_i+n_j)a_j
 +  {\rm h.\,c.} \bigr] \Bigr\} \ .
\nonumber
\end{eqnarray}
The frequency-independent $1/S$ correction to the magnon energy
due to $\hat{\cal H}_4$ is most easily found by performing
the Hartree-Fock decoupling  in Eq.~(\ref{H4}) with
the following mean-field averages  \cite{Oguchi60}
\begin{eqnarray}
&& n = \langle a_i^{\dagger} a_i \rangle =  \sum_{\bf k} v_{\bf k}^2\,,
\ \ m=\langle a_i^{\dagger} a_j \rangle =\sum_{\bf k}\gamma_{\bf k} v_{\bf k}^2\,, \\
&& \delta = \langle a_i^2 \rangle = \sum_{\bf k}  u_{\bf k} v_{\bf k}\,,
\ \ \Delta =  \langle a_i a_j \rangle =\sum_{\bf k} \gamma_{\bf k} u_{\bf k} v_{\bf k} \ .
\nonumber
\end{eqnarray}
The frequency-dependent corrections from quartic anharmonicities
appear only in the next  order  in $1/S$ and are not considered here.

The role of the cubic anharmonicity is twofold. First,
the Hartree-Fock decoupling in Eq.~(\ref{H3}) yields quantum correction
to the canting angle: \cite{Zhitomirsky98}
\begin{equation}
\sin \bar{\theta} = \sin{\theta} \Bigl[ 1 +
                    \frac{1}{S}(n-m-\Delta) \Bigr] \ .
\end{equation}
Substituting $\sin\bar{\theta}$ back
into Eq.~(\ref{H2}) one obtains another frequency-independent
correction to the spectrum. The combined Hartree-Fock correction, which includes the
above two  contributions, reads as
\begin{eqnarray}
\delta\epsilon^{(1)}_{\bf k} & = &\frac{4J}{\omega_{\bf k}}
\Bigl[\Delta-n+\sin^2\!\theta(\Delta+m)(1 - 2\gamma_{\bf k}\cos^2\!\theta)
\label{deltaE1} \\
&+ & \gamma_{\bf k}^2\bigl\{n-\Delta-m +\cos 2\theta\,(\Delta\sin^2\!\theta+m\cos^2\!\theta)
\bigr\}\Bigr]
\nonumber
\end{eqnarray}
with  $\omega_{\bf k}$ being the dimensionless magnon energy, see Eq.~(\ref{epsilon}).

The second contribution is due to the remaining fluctuating terms  in $\hat{\cal H}_3$ 
described by three-magnon processes:
\begin{eqnarray}
\hat{V}_{3}^{(1)} &=& \frac{1}{2!} \sum_{{\bf k},{\bf q}}
\Phi_1({\bf k},{\bf q})\: \Bigl[ b^\dagger_{\bf k-q+Q}
b^\dagger_{\bf q} b_{\bf k}  + {\rm h.\,c.}\Bigr],
\label{V31} \\
\hat{V}_{3}^{(2)} &=& \frac{1}{3!}  \sum_{{\bf k},{\bf q}}
\Phi_2({\bf k},{\bf q})\: \Bigl[
b^\dagger_{\bf k}  b^\dagger_{\bf q}  b^\dagger_{\bf Q-k-q}
+ {\rm h.\,c.} \Bigr].
\label{V32}
\end{eqnarray}
Decay and source  vertices are given explicitly by
\begin{eqnarray}
\Phi_{1,2}({\bf k},{\bf q}) & = & -H \cos{\theta} \;
\widetilde{\Phi}_{1,2}({\bf k},{\bf q})/\sqrt{2SN}\ ,
\label{Phi12} \\
\widetilde{\Phi}_1  (\mathbf{k},\mathbf{q}) & = &
\gamma_{\mathbf{k}} ( u_{\mathbf{k}}+v_{\mathbf{k}})
(u_{\mathbf{q}} v_{\mathbf{k-q+Q}}\! + v_{\mathbf{q}} u_{\mathbf{k-q+Q}})
\nonumber \\
& + & \gamma_{\mathbf{q}} (u_{\mathbf{q}}+v_{\mathbf{q}})
(u_{\mathbf{k}}u_{\mathbf{k-q+Q}}\! + v_{\mathbf{k}}v_{\mathbf{k-q+Q}})
\nonumber \\
& + & \gamma_{\mathbf{k-q+Q}}(u_{\mathbf{k-q+Q}}\!+\!v_{\mathbf{k-q+Q}})
(u_{\mathbf{k}}u_{\mathbf{q}}\!+\! v_{\mathbf{k}}v_{\mathbf{q}}),
\nonumber \\
\widetilde{\Phi}_2  (\mathbf{k},\mathbf{q}) & = &
\gamma_{\mathbf{k}} ( u_{\mathbf{k}}+v_{\mathbf{k}})
(u_{\mathbf{q}} v_{\mathbf{k+q-Q}}\! + v_{\mathbf{q}} u_{\mathbf{k+q-Q}})
\nonumber \\
& + & \gamma_{\mathbf{q}} (u_{\mathbf{q}}+v_{\mathbf{q}})
(u_{\mathbf{k}}v_{\mathbf{k+q-Q}}\! + v_{\mathbf{k}}u_{\mathbf{k+q-Q}})
\nonumber \\
& + & \gamma_{\mathbf{k+q-Q}}(u_{\mathbf{k+q-Q}}\!+\!v_{\mathbf{k+q-Q}})
(u_{\mathbf{k}}v_{\mathbf{q}}\! + \!v_{\mathbf{k}}u_{\mathbf{q}}).
\nonumber
\end{eqnarray}

Next, we define the bare magnon Green's function
\begin{equation}
G^{-1}_0({\bf k},\omega) =  \omega -\epsilon_{\bf k} + i 0
\label{G0kw}
\end{equation}
and calculate the second-order self-energy correction generated by the cubic terms,
which  are represented by two diagrams in  Fig.~\ref{diagrams}:
\begin{eqnarray}
\Sigma_1({\bf k},\omega) & = & \frac{1}{2} \sum_{\bf q}
\frac{|\Phi_1({\bf k},{\bf q})|^2} {\omega -\epsilon_{\bf q}-
\epsilon_{\bf k-q+Q} + i0} \ ,
\label{Sigma1} \\
\Sigma_2({\bf k},\omega) & = & -\frac{1}{2} \sum_{\bf q}
\frac{|\Phi_2({\bf k},{\bf q})|^2} {\omega +\epsilon_{\bf q}+
\epsilon_{\bf k+q-Q} - i0} \ .
\label{Sigma2}
\end{eqnarray}
In the leading $1/S$ order quantum correction is given by
the on-shell self-energy:
\begin{eqnarray}
&& \delta \epsilon^{(2)}_{\bf k} \equiv \Sigma({\bf k},\epsilon_{\bf k}) =  -4 J \sin^2\!\theta\, \cos^2\!\theta
\label{deltaE2} \\
&& \phantom{\epsilon^{(2)}}  \times
\sum_{\mathbf{q}} \Bigl[\frac{|\tilde{\Phi}_{1}(\mathbf{k,q)}|^2}
{\omega_{\bf q}\! + \omega_{\bf k-q+Q}\! - \omega_{\bf k}}
+ \frac{|\tilde{\Phi}_{2}(\mathbf{ k,q})|^2}
{\omega_{\bf k} \! + \omega_{\bf q}\! + \omega_{\bf k+q-Q}} \Bigr]
\nonumber
\end{eqnarray}
with the net renormalization for the excitation energy:
\begin{equation}
\bar{\epsilon}_{\bf k}  = \epsilon_{\bf k} + \delta\epsilon^{(1)}_{\bf k} + \delta\epsilon^{(2)}_{\bf k} \ .
\label{Eren}
\end{equation}
The magnon decay rate in the $1/S$ order is given by
\begin{equation}
\Gamma_{\bf k} =
\frac{\pi}{2}
\sum_{\mathbf{q}} |\Phi_{1}(\mathbf{k,q)}|^2
\delta\bigl( \epsilon_{\bf k} - \epsilon_{\bf q} - \epsilon_{\bf k-q+Q}\bigr) \ .
\label{Gk}
\end{equation}
Note, that  the above Born approximation yields a spin-independent value of the
decay rate $\Gamma_{\bf k}=O(J)$.

\begin{figure}[t]
\includegraphics[width=0.85\columnwidth]{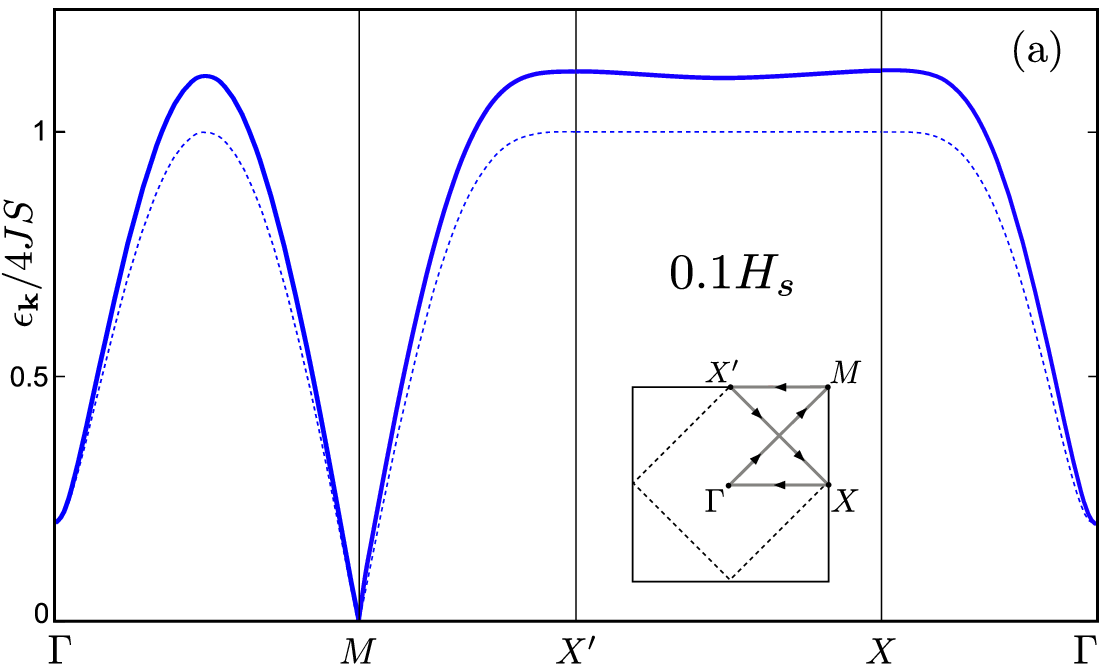}
\vspace*{5mm}

\includegraphics[width=0.85\columnwidth]{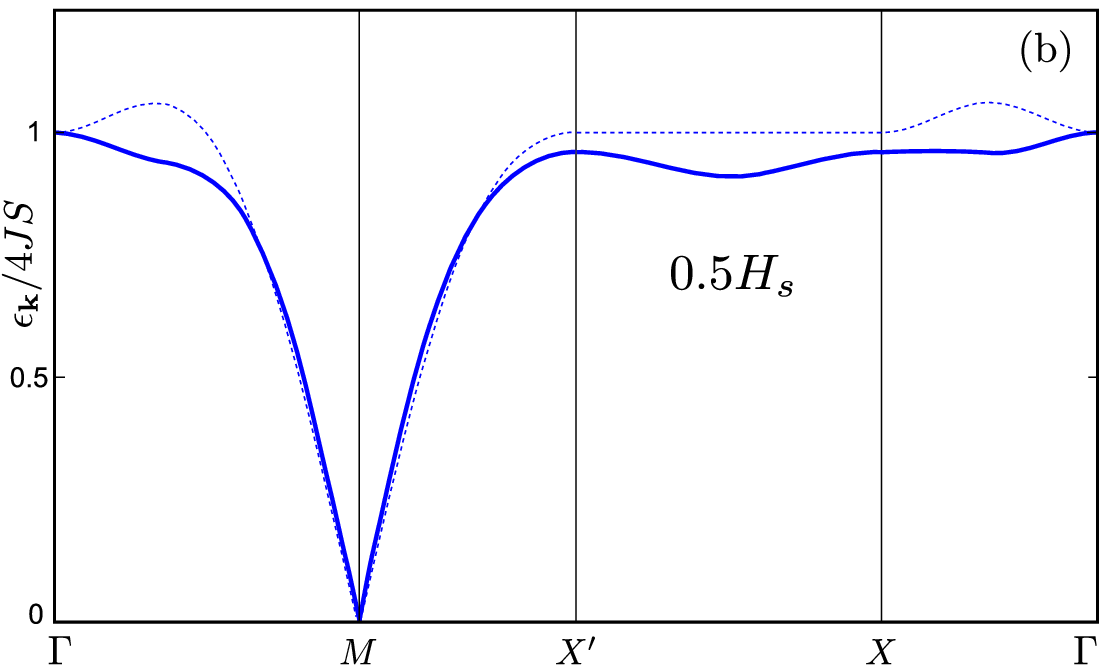}
\caption{(Color online)
Magnon energy of the spin-1/2 square-lattice antiferromagnet below
the decay threshold. Dashed lines
show the harmonic spin-wave energy $\epsilon_{\bf k}$, solid lines
represent the spectrum with the lowest-order quantum correction. The 
inset in (a) defines the chosen path in the Brillouin zone.
}
\label{dispersion_low}
\end{figure}

\begin{figure}[t]
\includegraphics[width=0.85\columnwidth]{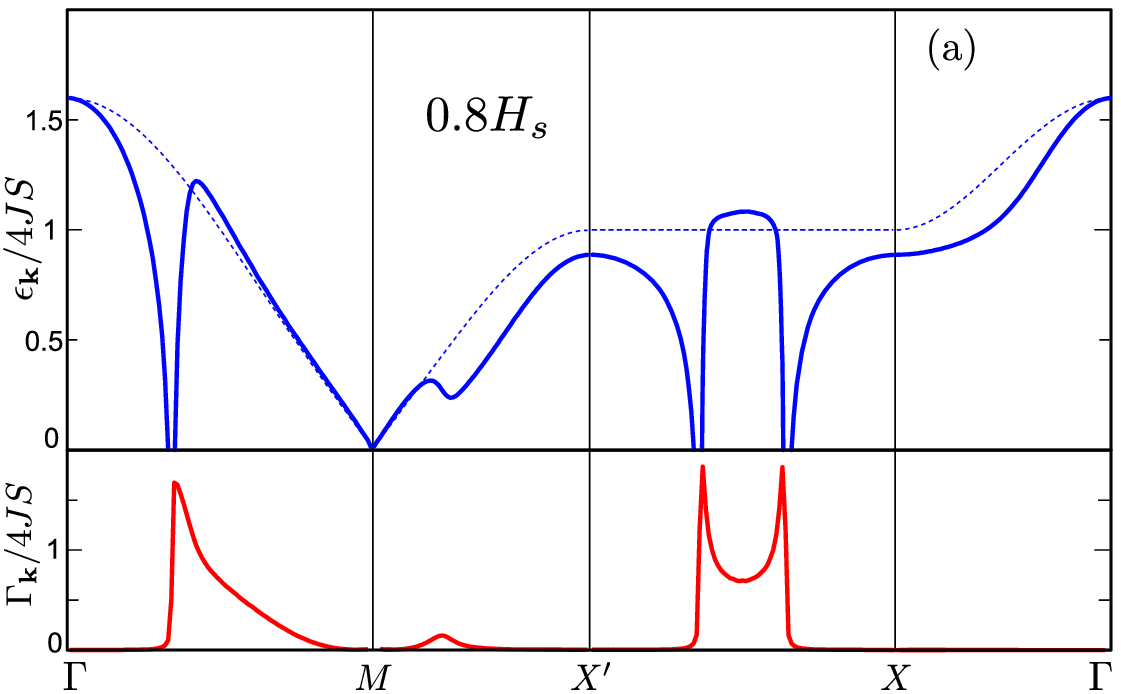}
\vspace*{5mm}

\includegraphics[width=0.85\columnwidth]{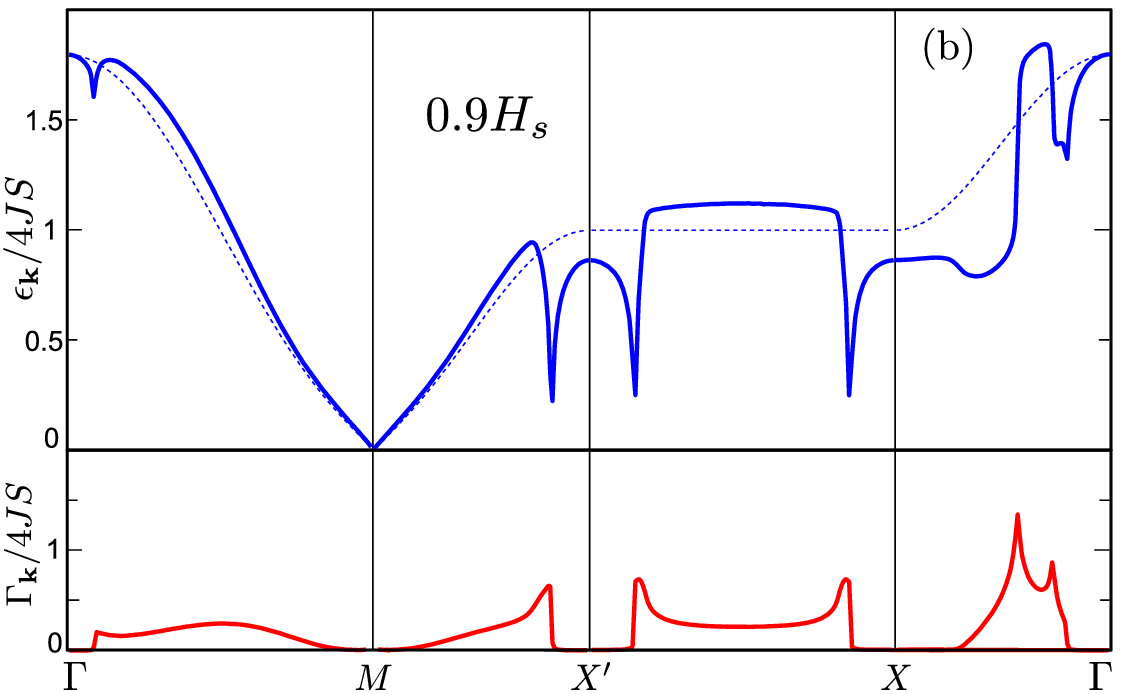}
\caption{(Color online)
Magnon energy and decay rate of the spin-1/2 square-lattice antiferromagnet
above the decay threshold. Upper panels:
dashed lines show the harmonic spin-wave energy
$\epsilon_{\bf k}$, solid lines represent the spectrum with
the lowest-order quantum correction $\bar{\epsilon}_{\bf k}$.
Lower panels: the decay rate $\Gamma_{\mathbf{k}}$ in
the Born approximation.
}
\label{dispersion_high}
\end{figure}

Although, our focus in Sec.~IV is on the large-$S$ model, we now present
numerical data for the spectrum renormalization (\ref{Eren})
in the $S=1/2$ case. Similar results for the other values
of spin can be obtained by rescaling quantum corrections in Figs.~\ref{dispersion_low}
and \ref{dispersion_high} by the  factor $1/(2S)$.
For moderate magnetic fields, $H\alt 0.6H_s$, the magnitude of
the correction to the magnon energy
is comparable to that in zero field. \cite{Zhitomirsky98b,Zhitomirsky99}
Numerical results for $\bar{\epsilon}_{\bf k}$ along the symmetry
directions in the Brillouin zone are shown in Fig.~\ref{dispersion_low}
for two values of external field.  Contrary to the $H=0$  case,
the renormalization is momentum-dependent
with increasing deviations from the harmonic theory as the field
approaches the decay threshold $H^*\approx 0.75 H_s$.
Specifically, there is a sizable dispersion
along the magnetic zone boundary $X^{\prime}X$,
which further increases in fields above one-half of the
saturation field. This is precisely the field regime
where the hybridization between one- and two-magnon states
grows substantially.

The lowest-order spin-wave correction changes  dramatically at higher
fields. The renormalized magnon energy $\bar{\epsilon}_{\bf k}$ exhibits
peculiar singularities in the form of jumps and spikes, see
Fig.~\ref{dispersion_high}. These anomalies signify a breakdown of the
perturbative spin-wave expansion since the $1/S$ correction appears to be divergent
for certain momenta. Such an outcome is rather surprising
because the expansion parameter $\langle a^\dagger_ia_i\rangle/(2S)$
decreases monotonically with the field and vanishes at $H=H_s$.
In particular, the $1/S$ spin-wave series for the SAFM
 converges rapidly for the ground-state energy and uniform magnetization,\cite{Zhitomirsky98} which agree very well 
with the numerical results. \cite{Luscher09}

The observed anomalies in the dynamical properties are clearly related to
zero-temperature magnon decays: every singularity in $\bar{\epsilon}_{\bf k}$
is accompanied by a jump  or by a peak in the decay rate
$\Gamma_{\bf k}$. \cite{Zhitomirsky99,Chernyshev06}
The typical magnitude of $\Gamma_{\bf k}$ for high-energy magnons
is comparable to or even exceeds $\epsilon_{\bf k}$, which
means that the dynamical response in certain parts of the Brillouin
zone is determined predominantly by the two-magnon continuum.
A detailed consideration of the origin of the decay anomalies
is provided in the next Section, while the asymptotic expressions for
$\Gamma_{{\bf k}\to{\bf Q}}$ are derived in Appendix A.

\section{Kinematics of magnon decays}
\label{kinematics}

The problem of spontaneous (zero-temperature) quasiparticle decays
generated by cubic anharmonicities has a rather long history. The phenomenon is
comprehensively documented for phonons in crystals \cite{Klemens67}
and for  excitations in superfluid $^4$He. \cite{Maris77,Pitaevskii}
Similar effects in quantum magnets have started to attract attention only recently.
\cite{Affleck92,Zhitomirsky99,Chernyshev06,Stone06,Masuda06,Veillette05,Zhitomirsky06}
Appearance  of spontaneous two-magnon decays for the SAFM  is controlled
by the energy conservation law:
\begin{equation}
\epsilon_{\bf k} = \epsilon_{\bf q} + \epsilon_{\bf k-q+Q} \ .
\label{energy_conserv}
\end{equation}
The ordering wave-vector $\bf Q$ enters the momentum conservation
condition because of the staggered canting of local moments
(\ref{eq:frame}).	
If the energy conservation is satisfied only for a trivial solution
${\bf q}={\bf Q}$, then magnon with the momentum $\bf k$ remains
stable.  The decay rate $\Gamma_{\bf k}$ becomes finite if there are
non-trivial solutions of Eq.~(\ref{energy_conserv}).
A less obvious but not less generic outcome of the mixing between one- and two-particle
states is the transfer of some of the Van Hove singularities
from the two-particle density of states onto the single-particle spectrum
producing various non-analyticities in the latter.\cite{Chernyshev06}
Thus, studying the kinematics of two-magnon decays
for a given dispersion $\epsilon_{\bf k}$
one should consider two related problems:
(i)~what is the {\it decay region} in the momentum space where
excitations are  unstable,
and (ii) where do the renormalization corrections exhibit
singular behavior.
In the present case, we are also interested in the
evolution of both the decay region and the singularities
as a function of external magnetic field.

Let us begin with the decay region. At the  boundary of such a region
the single-magnon branch crosses with the bottom of the two-magnon
continuum. Therefore, the boundary
can be determined by solving the system of two equations:
Eq.~(\ref{energy_conserv}) and
the extremum condition imposed on its r.h.s.:
\begin{equation}
{\bf v}_{\bf q} = {\bf v}_{\bf k-q+Q}
\ ,
\label{velo_conserv}
\end{equation}
where ${\bf v}_{\bf k} = \nabla_{\bf k}\epsilon_{\bf k}$
is a magnon velocity.
Generally, one can envisage
several types of solutions for the decay threshold: \cite{Pitaevskii,Chernyshev06} \\
(i) decay threshold with the emission of an acoustic magnon
$\bf q\rightarrow Q$,
determined by the condition
\begin{equation}
|{\bf v}_{\bf k}| = |{\bf v}_{\bf Q}| \equiv c \ ,
\label{goldstone}
\end{equation}
(ii) decay threshold with the emission of two magnons with equal momenta
${\bf q} =\frac{1}{2}({\bf k}+{\bf Q})$
found by solving
\begin{equation}
\epsilon_{\bf k} = 2\epsilon_{({\bf k}+{\bf Q})/2} \ ,
\label{equiv}
\end{equation}
and (iii) decay threshold with the emission of two nonequivalent
magnons found from a direct numerical solution of Eqs.~(\ref{energy_conserv})
and (\ref{velo_conserv}). Then, the decay region is the union of
regions obtained for all three decay channels.

\begin{figure}[t]
\begin{center}
\includegraphics[width=0.75\columnwidth]{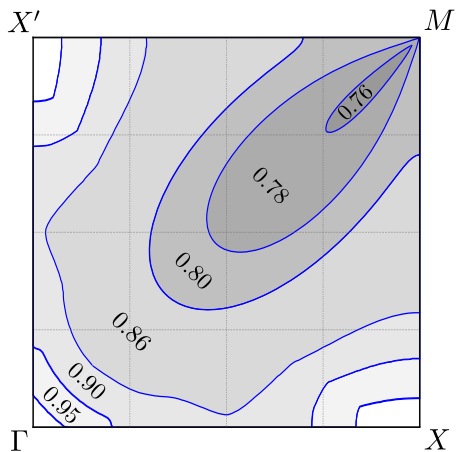}
\end{center}
\caption{(Color online) Field evolution of the decay region of the SAFM in the Born
approximation. Numbers correspond to values of $H/H_s$.}
\label{decayregion}
\end{figure}

Magnons in the square-lattice antiferromagnet remain stable up to the threshold
field $H^*\approx 0.75H_s$. \cite{Zhitomirsky99}
Numerical results for the  decay regions
in higher magnetic fields are shown in Fig.~\ref{decayregion}.
In fields slightly above $H^*$, the instability first affects
the acoustic branch producing
a cigar-shaped decay region oriented along the $\Gamma M$ diagonal
in the Brillouin zone.
Expanding energy of the acoustic magnon we obtain
\begin{eqnarray}
&& \epsilon_{\bf Q+k} \approx c k + \alpha k^3\ , \qquad
c = 2\sqrt{2}JS\cos\theta \ ,
\nonumber
\\[0.7mm]
&&
\alpha = \frac{c}{16}
\Bigl(\tan^2{\theta}-\frac{9+\cos{4\varphi}}{6} \Bigr) \ ,
\label{E_acoust}
\end{eqnarray}
where $c$ is the  acoustic magnon velocity and $\alpha$ is
the coefficient of the leading nonlinearity in the dispersion,
which depends on the momentum direction via the azimuthal
angle $\varphi$. Then, using $\sin{\theta}=H/H_s$, one can easily find
that $\alpha$ changes its sign from negative in low fields
to positive in high fields. Such a change in convexity
of the acoustic branch is generally known to result in the kinematic instability
with respect to two-particle decays, \cite{Maris77,Pitaevskii}
see also discussion in Appendix A.
The sign of $\alpha$ changes first
for $\varphi=\pi/4$  at the threshold field
\begin{equation}
H^* = \frac{2}{\sqrt{7}} \ H_{s} \approx 0.7559 \,H_s\ .
\label{H*}
\end{equation}
The above consideration is based on the harmonic expression for $\epsilon_{\bf k}$.
Quantum corrections to the magnon dispersion may renormalize
the value (\ref{H*}), although  the corresponding effect is expected to be small
for $S>1/2$.

The described change in the curvature of the Goldstone branch with increasing
field is a rather general phenomenon. The acoustic mode
changes convexity at the same field (\ref{H*}) in the cubic-lattice
Heisenberg antiferromagnet as well as in the stacked square-lattice
model for any value of the  antiferromagnetic coupling between layers.
Moreover, the spin-wave velocity vanishes at $H=H_s$ for
all Heisenberg antiferromagnets. In this critical field the
magnon branch is parabolic at low energies: $\epsilon_{\bf Q+k}\propto k^2$.
By the continuity argument, the spectrum preserves its positive
curvature for a certain range of magnetic fields in the ordered phase,
where  the asymptotic form (\ref{E_acoust}) holds.
This, in turn, implies that the two-particle decays are allowed at least for
${\bf k}\rightarrow{\bf Q}$.  A well-known example of such a behavior
is provided by the hard-core Bose gas. \cite{Beliaev58}
More generally, 
low-energy excitations in the ordered phase near $XY$ quantum critical
point are always unstable with respect to two-particle decays. 
Their decay rate has a universal form $\Gamma_{\bf k}\propto k^{2D-1}$,
in agreement with previous results  for $D = 2$ and 3. \cite{Chernyshev06,Kreisel08}
A more detailed discussion of the decay rate of low-energy magnons in the SAFM
is provided in Appendix A.

Magnetic systems with the $O(2)$ rotational symmetry about the field direction
belong to the same $XY$ universality class as the Heisenberg SAFM in a field.
An analysis of the kinematic conditions for magnon decays in the $XXZ$
square-lattice antiferromagnet in a field parallel to the $\hat{\bf z}$-axis
is briefly summarized in Appendix B. Another common example is given by
quantum spin systems with the singlet ground state and
gapped triplet excitations. The field-induced ordering in such systems
has been intensively studied both experimentally
\cite{Nikuni00,Jaime04,Regnault06} and theoretically.
 \cite{Affleck91,Affleck92,Mila98,Giamarchi99}
The mechanism of spontaneous magnon decays discussed here applies
equally to these magnets
in the vicinity of the triplet condensation field $H_c$ in the ordered phase because of the duality
between $H_c$ and $H_s$. \cite{Mila98}

For the Heisenberg SAFM,
the decay region  quickly spreads out across the Brillouin zone
as the applied field increases above $H^*$
and at $H>0.9H_s$ most of the magnons become unstable already
in the Born approximation,
see Fig.~\ref{decayregion}. Up to $H\approx 0.85H_s$ the boundary
of the decay region is entirely determined by
the decay into a pair of magnons with equal momenta (\ref{equiv}).
At higher fields $H \agt 0.85 H_s$ the decay channel with emission
of an acoustic magnon (\ref{goldstone}) starts to prevail in some parts of
the Brillouin zone producing a more complicated shape of the decay region,
see Fig.~\ref{decayregion}.

We now turn to the anomalous features in the magnon spectrum.
A close inspection of Fig.~\ref{dispersion_high} reveals two distinct
types of singularities in it.
The first type is characterized by a dip in $\bar{\epsilon}_{\bf k}$
and by a jump in $\Gamma_{\bf k}$,
while the anomaly of the second type consists in a jump
in $\bar{\epsilon}_{\bf k}$ accompanied by a peak in
$\Gamma_{\bf k}$.
The imaginary part of the magnon self-energy
is related to the two-magnon density of states via Eq.~(\ref{Gk}).
Hence, in 2D the decay rate
$\Gamma_{\bf k}$  exhibits a finite jump upon entering
the continuum in accordance with the corresponding Van Hove
singularity in the two-particle density of states:
\cite{Zhitomirsky99}
\begin{equation}
\Gamma_{\bf k} \simeq \Theta(\Delta k)\ , \qquad
\textrm{Re}\,\Sigma_{\bf k}(\epsilon_{\bf k})\simeq
\ln\frac{\Lambda}{|\Delta k|} \ ,
\label{log1}
\end{equation}
where $\Lambda$ is a cutoff parameter. The associated singularity in $\textrm{Re}\,\Sigma_{\bf k}(\epsilon_{\bf k})$
follows from the Kramers-Kronig relation.

\begin{figure}[t]
\includegraphics[width=0.75\columnwidth]{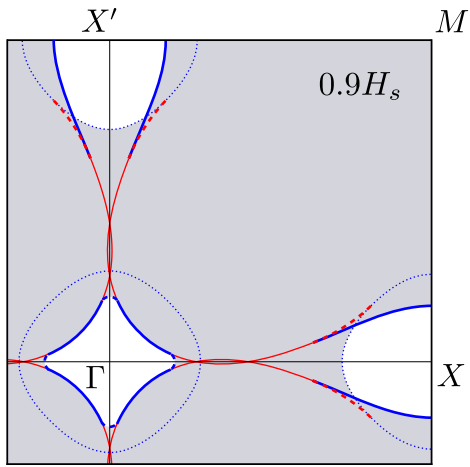}
\caption{(Color online)
Decay region and associated singularities for $H = 0.9H_s$.
Line's type represents  the decay channel:
solid is  for a pair of magnons with equal momenta,
dashed is for a pair of magnons with different momenta, and
dotted is for emission of an acoustic magnon.
Line's color indicates the type of singularity in
the continuum: blue is for the  minimum  and
red is for the saddle-point.  }
\label{decaymodes90}
\end{figure}

The second type of anomaly corresponds to an intersection
of the one-magnon branch with the
saddle-point
Van Hove singularity inside the continuum:  \cite{Chernyshev06}
\begin{equation}
\Gamma_{\bf k}
\simeq  \ln\frac{\Lambda}{|\Delta k|} \ , \qquad
\textrm{Re}\,\Sigma_{\bf k}(\epsilon_{\bf k})\simeq
\textrm{sign}(\Delta k)\ .
\label{log2}
\end{equation}
The boundary line given  by Eq.~(\ref{equiv}) always corresponds to a local
extremum and when it enters the interior of the decay region its type changes
from a local minimum to a saddle point. The details of such a behavior are
illustrated in Fig.~\ref{decaymodes90}. Here, solid lines denote
singularity contours for the decay into equivalent magnons,
dashed lines indicate the decay threshold into a pair of magnons with
different momenta, and dotted lines represent the decay threshold
for emission of an acoustic magnon. The colors are used
to distinguish whether the extremum corresponds to a local
minimum (blue) or to a  saddle-point (red) in the continuum.
The same singularity contour may
correspond to a minimum in one part of the Brillouin zone  and to a saddle-point
in the other. This determines, in turn, whether the logarithmic singularity
occurs in the real or imaginary part of the self-energy.
Further enhancement of the magnon decay rate may occur in the vicinity
of the  intersection of two saddle-point lines, an example which
can be found in Fig.~\ref{decaymodes90}
along the $\Gamma X$ line.

We finish this section with a brief discussion of higher-order magnon decays.
Similar to the consideration of the cubic nonlinearities in Sec.~II, the quartic 
terms in Eq.~(\ref{H4}), treated beyond the Hartree-Fock approximation, produce 
decays in the three-particle channel: $b^\dagger_{\bf q} 
b^\dagger_{\bf p} b^\dagger_{\bf k-q-p} b_{\bf k}$. 
In addition, higher-order bosonic terms omitted in Sec.~II 
should open decays of one magnon into an arbitrary large number of magnons. 
Nevertheless, the kinematic analysis above 
still predicts the correct threshold field for magnon decays.
This is because all higher-order $n$-particle decays with $n\geq 3$ are 
energetically forbidden if no two-magnon decays are allowed in the whole
Brillouin zone. \cite{Harris71}
In our case, this means that magnons in the SAFM remain 
completely stable up to the two-magnon threshold field $H^*$.
Once two-magnon decays become possible at $H>H^*$, the
$n$-magnon decays are also kinematically allowed. 
Generally, we expect the corresponding decay rates 
to be significantly smaller than for the two-magnon decays because of
their higher $1/S$ order. Moreover, the multi-magnon decays should not 
produce any singularities of the type (\ref{log1}) and (\ref{log2})
due to higher-dimensional integration in corresponding analogs of Eq.~(\ref{Gk}).

\section{Self-consistent theory of magnon decays}
\label{selfconsistent}

\subsection{One-magnon Green's function}

\begin{figure}[b]
\includegraphics[width=0.9\columnwidth]{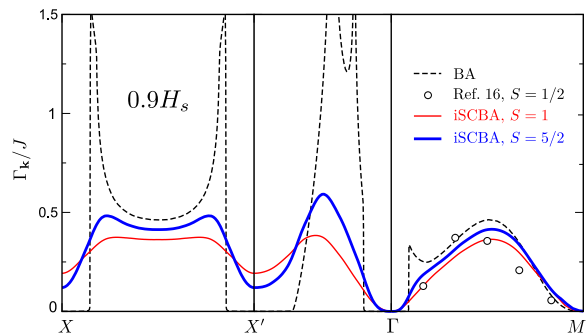}
\caption{(Color online)
Magnon decay rate $\Gamma_{\mathbf{k}}$ at $H=0.9H_s$
calculated in the Born approximation (dashed line) and in the iSCBA scheme
for $S=1$ (thin solid line, red) and for $S=5/2$ (bold solid line, blue).
Dots in $\Gamma M$ panel are $-{\rm Im}\Sigma_{\bf k}(\bar{\epsilon}_{\bf k})$
for $S=1/2$ from Ref.~\onlinecite{Zhitomirsky99}.
}
\label{scba_0.9}
\end{figure}

\begin{figure*}[tbh!]
\includegraphics[width=1.8\columnwidth]{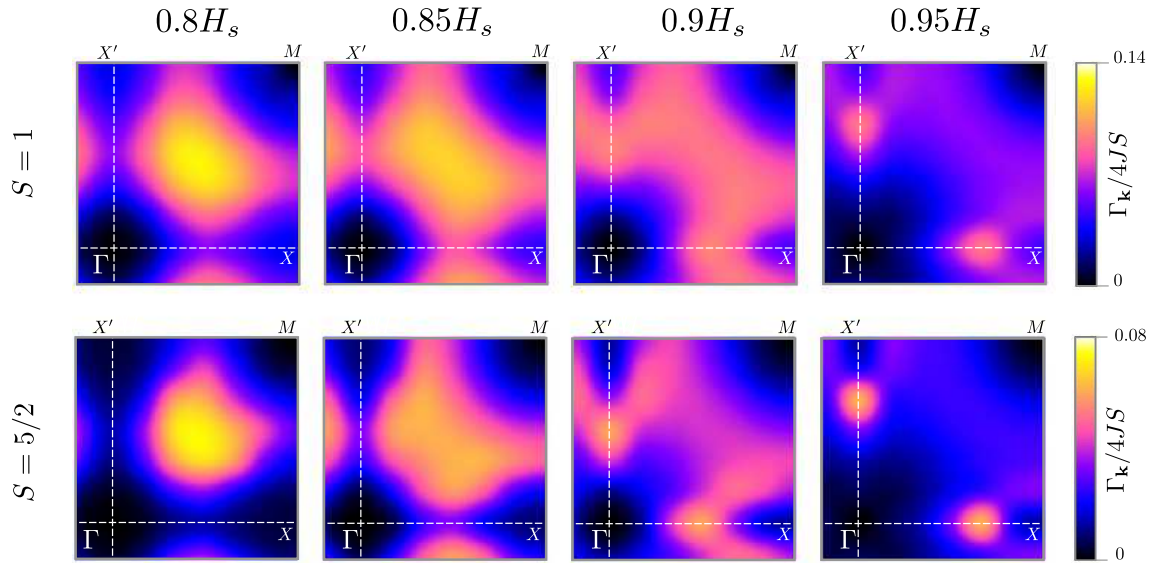}
\caption{(Color online)
Intensity maps of the magnon decay rate $\Gamma_{\mathbf{k}}$
calculated within the iSCBA  for several values of external magnetic field
and for the two values of the spin:  $S=1$ (upper panels) and
$S=5/2$ (lower panels).
}
\label{decay_rate}
\end{figure*}

Singular quantum corrections to the magnon spectrum obtained in the leading
$1/S$ order are generally regularized by higher-order processes.
Details of such a regularization depend on the specific shape of
$\epsilon_{\bf k}$.  An important aspect of the kinematics of the field-induced decays in SAFM is that the low-energy magnons created
in a decay process are unstable by themselves.
Qualitatively, the inverse life-time $\Gamma_{\bf q}$ of the decay products
cuts off the singularities in (\ref{log1}) and (\ref{log2}) by
changing
$$
|\Delta k| \rightarrow \sqrt{(\Delta k)^2 + (\Gamma_{\bf q}/c_{\bf q})^2}
\ ,
$$
where $c_{\bf q}$ can be expressed through the velocities of the decay products.
Accordingly, the self-consistent procedure,
which replaces the bare magnon Green's functions in the self-energy diagrams
in Fig.~\ref{diagrams} with the renormalized ones, should produce
non-singular quantum corrections. The main problem with such a
self-consistent Born approximation (SCBA) is that it opens
an unphysical gap for acoustic magnons in violation of the Goldstone theorem.
Here we avoid this problem by performing a restricted self-consistent
calculation, which takes into account the imaginary part of the magnon self-energy and 
neglects the real part of it. Such an approximation, which is 
referred to as iSCBA in the following, is expected to yield reasonable accuracy
for spins $S\geq 1$.
Indeed, in the low-field region  the nonsingular quantum corrections to
the magnon energy do not exceed 15--18\% in the case of $S=1/2$,
see Sec.~II.  For larger spins this correction is further reduced by a factor
$1/(2S)$ and already for $S\agt 1$ it corresponds only to a small
shift of the magnon energy.

In the simplest realization of the iSCBA we further neglect
the frequency dependence of  $\textrm{Im}\,\Sigma_{\bf k}(\omega)$
and impose the following form of the one-magnon Green's function:
\begin{equation}
G^{-1}({\bf k},\omega) =  \omega -\epsilon_{\bf k} + i \Gamma_{\bf k} \ .
\label{Green1}
\end{equation}
The magnon decay rate $\Gamma_{\bf k}$ is calculated self-consistently from
\begin{equation}
\Gamma_{\bf k}  =  -\textrm{Im}\,\Sigma({\bf k},\epsilon_{\bf k}) \ ,
\label{Gamma_Lorentz}
\end{equation}
with $\Sigma({\bf k},\epsilon_{\bf k})$ from Fig.~\ref{diagrams}.
Physically, the approximation (\ref{Green1}) amounts to assuming
the Lorentzian shape of the quasiparticle peak in the dynamical response.
Such a self-consistent scheme has been previously applied to the
problem of phonon broadening
in superfluid helium\cite{Sluckin74} and in quasicrystals.\cite{Kats05}
In the present case, with two types of cubic vertices the self-consistent equation
on $\Gamma_{\bf k}$ becomes
\begin{eqnarray}
\Gamma_{\bf k} & = &
\frac{1}{2}\sum_{\bf q} \biggl[\frac{|\Phi_1({\bf k},{\bf q})|^2
(\Gamma_{\bf q}+\Gamma_{\bf k-q+Q})}
{(\epsilon_{\bf k}-\epsilon_{\bf q}-\epsilon_{\bf k-q+Q})^2 +
(\Gamma_{\bf q}+\Gamma_{\bf k-q+Q})^2}
\nonumber \\
&  + &  \frac{|\Phi_2({\bf k},{\bf q})|^2
(\Gamma_{\bf q}+\Gamma_{\bf k+q-Q})}
{(\epsilon_{\bf k}+\epsilon_{\bf q}+\epsilon_{\bf k+q-Q})^2 +
(\Gamma_{\bf q}+\Gamma_{\bf k+q-Q})^2} \biggr].
\label{SCBA}
\end{eqnarray}

One important consequence of the self-consistent consideration
is that spontaneous decays are not restricted to the decay region
in Fig.~\ref{decayregion}, defined in the Born approximation. 
Since the finite decay rate of magnons inside the decay region
means an uncertainty in their energy, the energy conservation
condition (\ref{energy_conserv}) is relaxed and magnons
just outside the decay region are also allowed to decay.
Therefore, the decay threshold boundary is smeared
and the decay probability disappears gradually rather than in a step-like fashion.

Numerical solutions of Eq.~(\ref{SCBA}) have been obtained iteratively using two
different grids in the momentum space: a sparse mesh for the momentum $\bf k$
of $40\times 40$ points in the full Brillouin zone
and a much tighter grid for integration over the momentum $\bf q$ with up to
$1000\times 1000$ points. The values of $\Gamma_{\bf q}$ for the integrand were
obtained from $\Gamma_{\bf k}$ by a bicubic interpolation. 
To improve convergence at low energies we also impose the 
asymptotic form $\Gamma_{\bf Q+k} \sim k^3$ for small $k$, see Appendix~A.
Sufficient
accuracy is typically achieved after 5--8 iteration steps.
Note, that the second term on the r.h.s.\ of Eq.~(\ref{SCBA}) from
the source diagram in Fig.~\ref{diagrams} is always small and can be
safely neglected. 

Figure~\ref{scba_0.9}
compares $\Gamma_{\bf k}$ obtained in the Born approximation with the iSCBA
results for $S = 1$ and $S = 5/2$ at $H=0.9H_s$. One can see that the singular 
behavior of  $\Gamma_{\bf k}$ near the Van Hove singularities
is completely removed in the self-consistent calculation. The typical amplitude of
the decay rate in Fig.~\ref{scba_0.9} away from $\Gamma$ and $M$ points is
$\Gamma_{\bf k} \alt 0.5J$, which should be compared with the typical 
magnon energies $\epsilon_{\bf k}\sim 4JS$.
Another notable aspect of the self-consistent regularization is in 
the spin-dependence of the decay rate: $\Gamma_{\bf k}$ is larger 
for larger spins in most of the Brillouin zone. Such a dependence should not be
confused with the {\it relative} strength of the spectrum broadening 
$\Gamma_{\bf k}/\epsilon_{\bf k}\sim \Gamma_{\bf k}/JS$, which  is
still larger for smaller values of the spin. The decay rate approaches 
the spin-independent Born result $\Gamma^{(0)}_{\bf k}$ only in the limit 
$S\to\infty$. In particular, while 
$\Gamma_{\bf k}\simeq \Gamma^{(0)}_{\bf k}+ O(1/S)$
away from the singular points, one finds  $\Gamma_{\bf k}\sim \ln S$ 
in their vicinity.\cite{Chernyshev06}

Further details on the momentum distribution of the magnon decay rate in 
different fields are provided in the form of intensity maps for 
$\Gamma_{\bf k}/4JS$ in Fig.~\ref{decay_rate}.
Comparison of the upper ($S=1$) and the lower ($S=5/2$) row reveals
similar patterns in the momentum distribution of $\Gamma_{\bf k}$ for these two
values of spin. The intensity of the decays in Fig.~\ref{decay_rate} is 
determined by either the volume of the decays or by the proximity to 
the Van Hove singularities of the two-magnon continuum.
Specifically, in the lower fields $H\alt 0.9H_s$ the magnon damping
is most significant in the broad region around $(\pi/2,\pi/2)$ point.
This is a consequence of the large phase space volume
for the two-particle decays in this region.
In higher fields the maximums in the decay rate shift towards
the $\Gamma X$ and $\Gamma X'$ lines.
The enhancement of $\Gamma_{\bf k}$ here is due to
an intersection of the two singularity lines discussed in Sec.~III.
Overall, the maximum magnon decay rate in all fields does not exceed 
$\Gamma^{\rm max}_{\bf k}\sim 0.7$--$0.8J$.
Thus, we may conclude that the spin waves in large-$S$ quantum antiferromagnets 
above the decay threshold $H^*$ are damped but still well-defined quasiparticles.

The above conclusion seems to be at odds with the behavior of the spin-1/2 SAFM.
In this case, magnons were shown to be overdamped in most of
the Brillouin zone by both the spin-wave calculation, which used a different version of the SCBA,
\cite{Zhitomirsky99} and by the numerical QMC simulations.\cite{Olav08}
Such a difference in the effect of magnon interaction between $S=1/2$ and $S\geq 1$ models 
deserves a separate discussion.
We compare directly the imaginary part of the magnon self-energy at the quasiparticle pole,
$-{\rm Im}\Sigma_{\bf k}(\bar{\epsilon}_{\bf k})$, for the $S=1/2$ case obtained in
Ref.~\onlinecite{Zhitomirsky99} with $\Gamma_{\bf k}$ found in this work,
see the data points along $\Gamma M$ direction in Fig.~\ref{scba_0.9}. One can see that $S=1/2$
and $S\geq 1$ results are quantitatively close. Naively, this would mean that 
quasiparticles in the $S=1/2$ antiferromagnet should also be reasonably 
well-defined. However, this consideration neglects the spectral weight 
redistribution. The analytical calculation \cite{Zhitomirsky99}  and 
the QMC study  \cite{Olav08} have shown a significant non-Lorentzian 
broadening  of the spectral lines, which takes the form of a double-peak 
structure in the dynamical susceptibility. Therefore, 
the overdamping in the $S=1/2$ case is produced by two
effects: broadening of the quasiparticle peak and spectral weight redistribution. 
Note, that the latter effect was completely excluded in the recent hard-core boson study,\cite{Syromyat09} 
which explicitly neglected the  frequency dependence of the magnon self-energy. 
Such an approximation may have lead Ref.~\onlinecite{Syromyat09} to 
the conclusion of only weak-to-moderate damping of spin excitation in 
the $S=1/2$ SAFM in a field, 
in contrast with Refs.~\onlinecite{Zhitomirsky99} and \onlinecite{Olav08}.

Since the iSCBA developed here also neglects the spectral weight redistribution, one can question its validity on the same grounds.
To verify the validity of this approach
we have  checked the accuracy of the Lorentzian
approximation used in (\ref{Green1}) by keeping the full frequency dependence
of  $\textrm{Im}\,\Sigma_{\bf k}(\omega)$ and by utilizing general representation 
for the magnon Green's function
\begin{equation}
G({\bf k},\omega) = \int_{-\infty}^\infty \textrm{d} x \, \frac{A_{\bf k}(x)}
{\omega-x+i0} \ .
\end{equation}
The self-consistent equation is formulated for the spectral function $A_{\bf k}(\omega)$ 
excluding again the real part of the magnon self-energy. Numerical results (not shown) exhibit only a small asymmetry of magnon peaks and only for the lowest $S=1$ case. This justifies the use of the frequency-independent
iSCBA scheme for $S\geq 1$ antiferromagnets and confirms the perturbative role of 
damping for them once the singularities are regularized by a self-consistent calculation.

\subsection{Dynamical structure factor}
\label{dynamics}

The inelastic neutron scattering experiments probe the spin-spin
correlation function
\begin{equation}
\mathcal{S}^{\alpha\beta}({\bf k},\omega) =
\int_{-\infty}^{\infty}  \frac{\mbox{d}t}{2\pi}\,
\langle S^{\alpha}_{\bf k }(t) S^{\beta}_{-\bf k } \rangle\, e^{i\omega t}
\label{Salphabeta}
\end{equation}
defined in the global laboratory frame,
while the spin-wave calculations are conveniently performed
in the local rotating frame, see Sec.~\ref{formalism}.  First, we relate
 these two forms of the correlation function with the help of Eq.~(\ref{eq:frame}):
\begin{eqnarray}
\mathcal{S}^{x_0x_0}({\bf k},\omega)  & = &  \sin^2\!\theta \,
\mathcal{S}^{xx}({\bf k},\omega) +  \cos^2\!\theta\,
\mathcal{S}^{zz}({\bf k}-{\bf Q},\omega)\,,
\nonumber \\
\mathcal{S}^{z_0z_0}({\bf k},\omega)  & = &
\cos^2\!\theta\, \mathcal{S}^{xx}({\bf k}-{\bf Q},\omega) +
\sin^2\!\theta \, \mathcal{S}^{zz}({\bf k},\omega)\,,
\nonumber               \\
\mathcal{S}^{y_0y_0} ({\bf k},\omega) & = & \mathcal{S}^{yy}({\bf k},\omega) \ .
\label{SQW1}	
\end{eqnarray}
Here we have omitted the cross-terms $(\alpha\neq \beta)$ on the r.h.s., which can be shown to be small numerically.
Second, we relate the dynamical structure factor (\ref{SQW1}) to
the time-ordered  spin Green's function $\mathcal{G}^{\alpha\beta}({\bf k},t) =
-i \langle T S^{\alpha}_{\bf k}(t) S^{\beta}_{-\bf k} \rangle$
using the fluctuation-dissipation theorem at zero temperature:
\begin{equation}
\mathcal{S^{\alpha\beta}}({\bf k,\omega})
= -\frac{1}{\pi}\, \mbox{Im}\,\mathcal{G}^{\alpha\beta}({\bf k},\omega) \ .
\label{FluctDisp}
\end{equation}
Finally, we express spin  operators via the Holstein-Primakoff bosons
and, after some of algebra, obtain the following expressions
\begin{eqnarray}
S^{xx}({\bf k},\omega) & = & - S\Lambda_+\: (u_{\bf k } + v_{\bf k })^2
\; \mbox{Im}\,G({\bf k,\omega}) \ ,
\nonumber\\
S^{yy}({\bf k},\omega) & = & - S\Lambda_- \: (u_{\bf k } - v_{\bf k })^2
\; \mbox{Im}\,G({\bf k,\omega}) \ ,
 \label{SQW2} \\
S^{zz}({\bf k},\omega) & = & -  \frac{1}{N}\displaystyle{\sum_{\bf q}}\,
(u_{\bf k-q} v_{\bf q}\! +\! u_{\bf q} v_{\bf k-q})^2\nonumber \\
& & \times\, \mbox{Im}\,\left\{
 \int_{-\infty}^{\infty}  \mbox{d}x \;
G({\bf q},x)\,G({\bf k}-{\bf q},\omega - x)\right\}\,.
\nonumber
\end{eqnarray}
Here $\Lambda_{\pm} = 1 - (2n\pm\delta)/2S$ are the Hartree-Fock spin
reduction factors and $G({\bf k},\omega)$ is the magnon Green's function
from (\ref{Green1}).

In the equation above,  spin fluctuations are separated into transverse ($xx$, $yy$)
and longitudinal ($zz$) components relative to the local spin direction, which correspond
to scattering from states with odd and even number of magnons, respectively.\cite{Canali93}
An important effect of the spin canting on the dynamical structure factor is the
redistribution of the spectral weight over the two well-separated transverse modes,\cite{Heilmann81}
which overlap only at the magnetic zone boundary. As a result,
Eq.~(\ref{SQW1}) contains a mixture of two transverse contributions at momenta
${\bf k}$ and ${\bf k-Q}$.
The first mode corresponds to spin fluctuations perpendicular to the
magnetic field direction (in-plane mode),
while the second mode, shifted by the ordering
wave-vector $\bf{Q}$, corresponds to the fluctuations along the field direction
(out-of-plane mode).
In strong magnetic fields the in-plane mode is enhanced and dominates over
the out-of-plane fluctuations, which gradually disappear as $H\rightarrow H_s$.
Note also the existence of two distinct longitudinal multimagnon continua
associated with each of the transverse branches.

\begin{figure}[t]
\centerline{
\includegraphics[width=0.9\columnwidth]{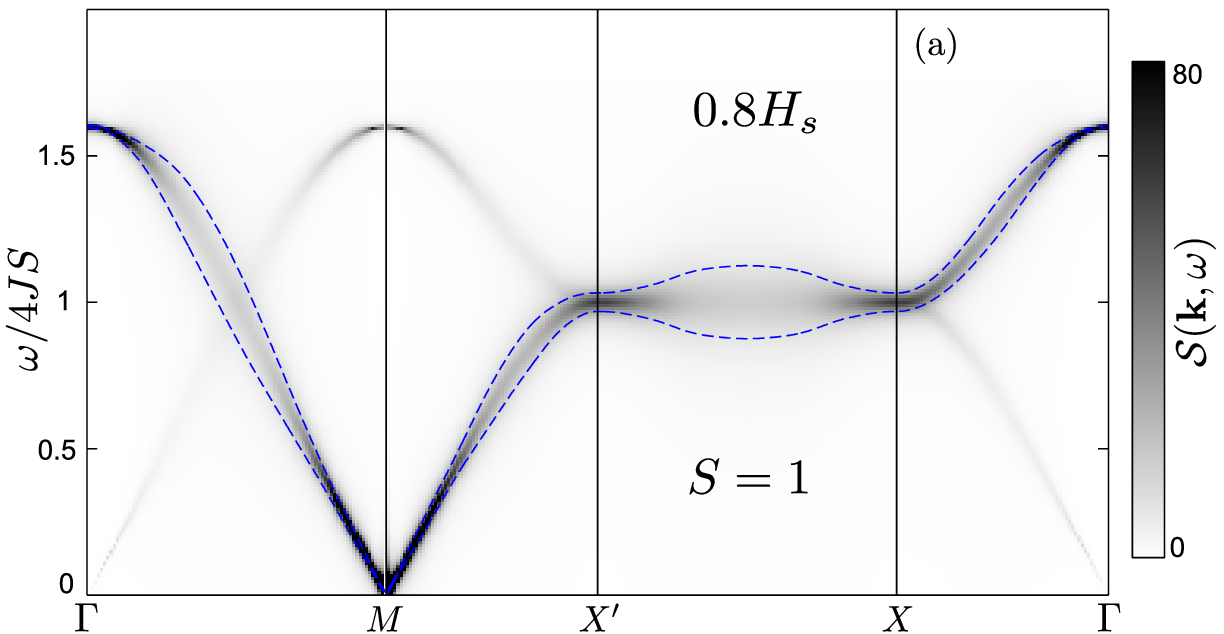}
}
\bigskip
\centerline{
\includegraphics[width=0.9\columnwidth]{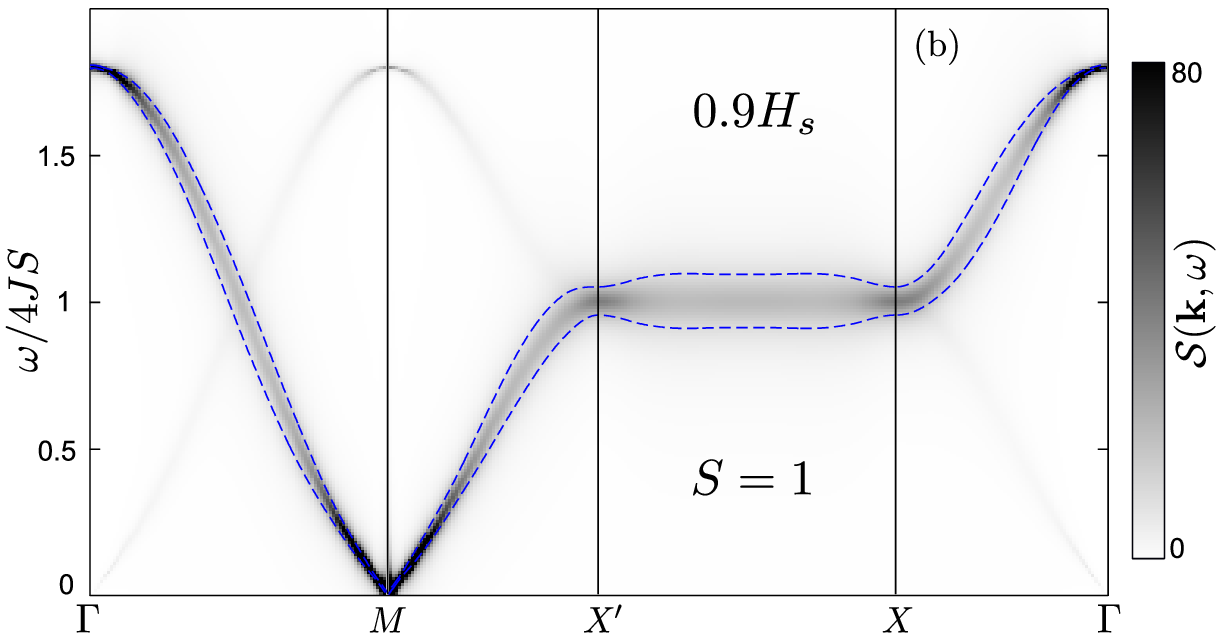}
}
\caption{(Color online) Transverse part of the dynamical structure
factor (gray scale) computed using the iSCBA for $S = 1$ and for the two values of external magnetic field. Dashed lines (blue)
enclose the region $\epsilon_{\bf k} \pm \Gamma_{\bf k}$
corresponding to the full-width at half maximum of the corresponding
quasiparticle peaks.
}
\label{Sqwcolorplot1}
\end{figure}

\begin{figure}[t]
\centerline{
\includegraphics[width=0.9\columnwidth]{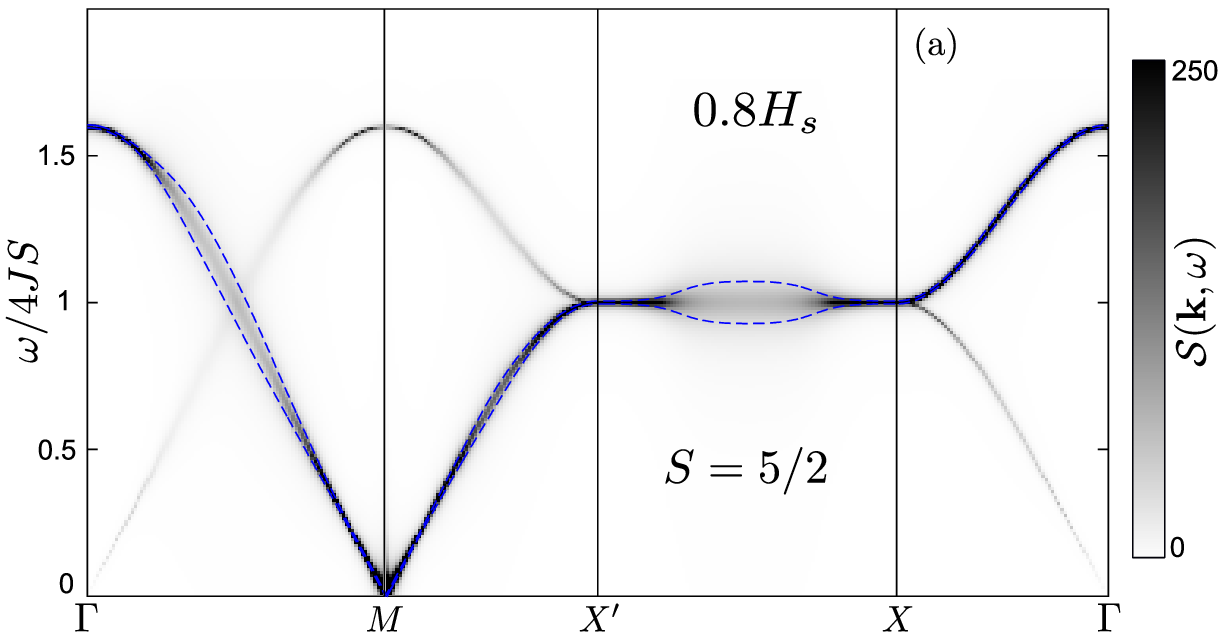}
}
\bigskip
\centerline{
\includegraphics[width=0.9\columnwidth]{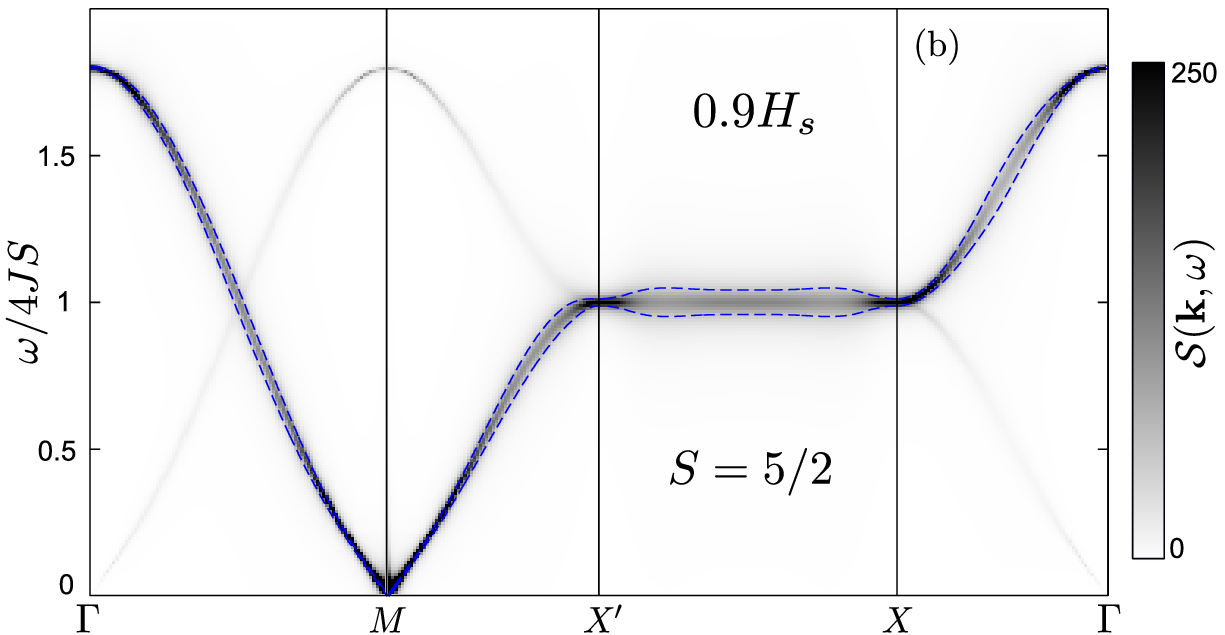}
}
\caption{(Color online) Same as Fig.~\ref{Sqwcolorplot1}
for $S = 5/2$.
}
\label{Sqwcolorplot52}
\end{figure}

Results summarized in Eqs.~(\ref{SQW1}) and (\ref{SQW2}) are quite general
and can be adapted to a variety of specific experimental situations.
As a particular example, the dynamical structure factor accessible
to unpolarized neutron scattering experiment with a vertical magnetic
field reads as
$\mathcal{S}({\bf k},\omega) = \mathcal{S}_{\perp}({\bf k},\omega) +
\mathcal{S}^{z_0z_0}({\bf k},\omega)$. The explicit expression for
$\mathcal{S}_{\perp} ({\bf k},\omega)$ depends on the experimental
details via a momentum-dependent polarization factor. In the following, such a
factor is deliberately ignored and the expression $\mathcal{S}({\bf k},\omega) =
\mathcal{S}^{x_0x_0} + \mathcal{S}^{y_0y_0}+ \mathcal{S}^{z_0z_0}$
is used for an illustration. The combined $\mathcal{S}({\bf k},\omega)$
computed using the self-consistent magnon Green's function obtained in
previous subsection is shown for $S = 1$ and $S=5/2$ antiferromagnets
in Figs.~\ref{Sqwcolorplot1} and \ref{Sqwcolorplot52}, respectively.
One can clearly see the presence of two transverse modes with the more intense
in-plane and the weaker out-of-plane mode.

A distinct fingerprint of spontaneous magnon decays as compared
to other possible damping mechanisms
is their characteristic dependence on the momentum and on the applied
magnetic field. In the field range $H^* \leq H \leq 0.9H_s$
the magnon damping is most significant along the magnetic zone boundary
$X^{\prime}X$ with the maximum at $(\pi/2,\pi/2)$. In higher magnetic fields
$H \geq 0.9 H_s$ the strongest damping occurs along the $\Gamma X$ line, where
the singularity-enhanced decays are most pronounced.
Our predictions for the former field regime are
in a qualitative agreement with the very recent inelastic neutron scattering results
of Masuda {\it et al.}\cite{Masuda10}\  for the spin-5/2 layered
square-lattice antiferromagnet Ba$_2$MnGe$_2$O$_7$.
However, the largest magnon linewidth measured experimentally
exceeds our theoretical estimate $2\Gamma_{\bf k} \simeq 1.5J$ for $H=0.82H_s$ by a factor of 2--3. The larger
experimental linewidth
may be related to the overlap of the two transverse modes along the magnetic zone boundary.
Therefore, we propose that the inelastic neutron scattering measurements at $H\geq 0.9H_s$ in the vicinity of
$(\pi/2,0)$ point where the two transverse modes are well separated
may provide a benchmark for the field-induced spontaneous magnon decays in the SAFM.

\section{Conclusions}

We have presented a detailed analysis of the field-induced decay dynamics
in the square-lattice antiferromagnet  within the framework of the spin-wave
theory. In magnetic fields exceeding the threshold field for two-particle decays, 
the $1/S$ quantum correction to the magnon spectrum exhibits singularities for
certain  momenta. These are related either to the decay thresholds
\cite{Zhitomirsky99} or to the saddle-point Van Hove singularities
in the two-magnon continuum.\cite{Chernyshev06}
Such singularities invalidate the usual $1/S$-expansion because
the resultant renormalized spectrum contains divergences.
The encountered problem is rather generic and must be common
to a variety of 2D models regardless of the on-site spin value.
For the systems with large spins the situation is especially aggravating as
the $1/S$ expansion turns from a       reliable approach to the one producing
unphysical divergences, which complicate any sensible comparison with experimental or
numerical data.

In the present work we have developed a self-consistent regularization scheme,
which is applicable to a variety of problems in the decay dynamics of
large-$S$ quantum antiferromagnets. The self-consistently calculated damping
in the SAFM is free from singularities and gives an upper limit on the
decay rate: $(\Gamma_{\bf k}/\epsilon_{\bf k})\alt 0.2/S$.
Overall, spin-waves in the SAFM with  $S\geq 1$ in magnetic field $H>H^*$ do acquire finite broadening, but remain well-defined quasiparticles. 
The regularized singularities also lead to a parametric enhancement of
the magnon damping  and produce a distinct  field- and momentum-dependence of
$\Gamma_{\bf k}$, which is illustrated for the $S=1$ and $S=5/2$
in Figs.~\ref{scba_0.9}-\ref{Sqwcolorplot52}.   These results can be used
for quantitative comparison in future experimental studies of the field-induced
spontaneous magnon decays.

For the spin-1/2 SAFM the situation appears to be somewhat more delicate.
Previous analytical\cite{Zhitomirsky99} and numerical\cite{Olav08} studies 
have predicted overdamped one-magnon excitations in a large part of 
the Brillouin zone.
The SCBA scheme used in Ref.~\onlinecite{Zhitomirsky99}
includes a self-consistent renormalization of only one inner magnon line in
the decay diagram of Fig.~\ref{diagrams} and is, in a sense, not as consistent
as the present approach. On the other hand, that approach does take into account
the real part of the spectrum renormalization and, as a consequence, 
the quasiparticle weight redistribution. Such an effect can be deemed small 
for larger spins, but it is more important for $S=1/2$ and is likely 
to contribute to further enhancement of the damping in this case. Since 
the imaginary part of the self-energy at the quasiparticle pole for $S=1/2$  
in Ref.~\onlinecite{Zhitomirsky99} correlates closely with the iSCBA results 
of the present work (Fig.~\ref{scba_0.9}), we believe that this is the correct 
explanation of the differences between $S=1/2$ and $S\geq 1$
cases. There are also qualitative and quantitative similarities of
the analytical \cite{Zhitomirsky99} and numerical results, \cite{Olav08}
in particular the double-peak structure in the spectral function.
While the QMC approach may involve its own uncertainties due to numerical 
interpolation from imaginary to real frequencies,\cite{Olav08} such 
a correspondence between  results of two very different methods is encouraging.
Further theoretical efforts may be needed to clarify completely 
the detailed behavior of the dynamical structure factor for the spin-1/2 SAFM.
Finally, performing inelastic neutron scattering measurements
on suitable spin-1/2 compounds would also be important.

\begin{acknowledgments}
Part of this work has been performed within the Advanced Study
Group Program on ``Unconventional Magnetism in High Fields'' at
the Max-Planck Institute for the Physics of Complex Systems.
The work  of one of us (A. L. C.) was supported by the DOE under grant
DE-FG02-04ER46174.
\end{acknowledgments}

\appendix

\section{Decay of low-energy magnons}

In the case of the Heisenberg
SAFM and other two-sublattice antiferromagnets in applied
magnetic field the low-energy spectrum consists of the
 single weakly nonlinear acoustic branch
(\ref{E_acoust}).
In the following consideration
all momenta are taken relative to the magnetic ordering wave-vector
 $\bf Q$: $\epsilon_{\bf Q+k}\rightarrow\epsilon_{\bf k}
\approx ck + \alpha k^3$.
First, we verify explicitly the energy conservation condition
for the two-particle decays (\ref{energy_conserv}). If the nonlinearity
of the spectrum is weak $\alpha k^3\ll ck$, the two magnons emitted in
a spontaneous decay process have their momenta $\bf q$ and
${\bf q}'={\bf k}-{\bf q}$ almost parallel to the direction of $\bf k$.
Then,
\begin{equation}
|{\bf k}-{\bf q}| \approx k-q + \frac{kq\varphi^2}{2(k-q)} \ ,
\end{equation}
where $\varphi$ is a small angle between $\bf q$ and $\bf k$, and
the energy conservation can be written as
\begin{eqnarray}
&& \epsilon_{\bf k} - \epsilon_{\bf q} - \epsilon_{\bf k-q} \approx
3\alpha kq(k-q) - \frac{ckq\varphi^2}{2(k-q)}
\label{Ekq}
\\
&& \mbox{} \quad = -\frac{ckq}{2(k-q)}\left(\varphi^2-\varphi_0^2\right) \ ,
\quad \varphi_0^2 = \frac{6\alpha}{c} (k-q)^2 \ .
\nonumber
\end{eqnarray}
One can see that the nontrivial solution
$\varphi=\varphi_0$ exists only for the
positive sign of the cubic nonlinearity $\alpha$.\cite{Maris77,Pitaevskii}

Next, we derive the asymptotic, $k,q\rightarrow 0$ form for the decay vertex
$\tilde\Phi_1({\bf k},{\bf q})$, see Eq.~(\ref{Phi12}). In order to obtain 
the correct expression one needs to retain three leading terms in the expansion of the Bogolyubov
coefficients $u_{\bf k}$ and $v_{\bf k}$ in small $k$. This yields
\begin{eqnarray}
\tilde{\Phi}_1 
& \approx & -\left(\frac{\cos\theta}{\sqrt{2}}
\right)^{1/2} \;\left[\frac{q+|{\bf k}-{\bf q}|-k}{\sqrt{kq(k-q)}} + \lambda \sqrt{kq(k-q)}\right] \ ,
\nonumber
\\
& & \lambda = \frac{3}{4\cos^2\theta} - \frac{3\alpha}{c} \ .
\end{eqnarray}
For momenta $\bf k$ and $\bf q$, which satisfy the energy conservation
condition (\ref{energy_conserv}), the decay vertex further simplifies into
\begin{equation}
\tilde{\Phi}_1({\bf k},{\bf q}) \approx  -\frac{3}{4\cos^2\theta}\,
\left(\frac{\cos\theta}{\sqrt{2}}\right)^{1/2}\,\sqrt{kq(k-q)} \ .
\label{Phi1simp}
\end{equation}
Substituting  (\ref{Ekq}) and (\ref{Phi1simp}) into the Born expression for 
the decay rate in  Eq.~(\ref{Gk}) one obtains
\begin{equation}
\Gamma_{\bf k} = \frac{3J}{16\pi} \, \tan^2\!\theta\;
\sqrt{\frac{c}{6\alpha}}\; k^3 \ .
\label{Gkasympt}
\end{equation}
The bare values of $c$ and $\alpha$ are given by Eq.~(\ref{E_acoust}),
although one may generally use renormalized parameters.  Note, that 
the asymptotic expression (\ref{Gkasympt}) 
differs from the result provided 
in Ref.~\onlinecite{Kreisel08}.
Therefore, we have verified that our expression for $\Gamma_{\bf k}$ in (\ref{Gkasympt}) 
agrees with the direct numerical integration of  Eq.~(\ref{Gk}) in the limit of small $k$.

As $H\to H_s$, the velocity of the acoustic mode decreases and the
condition of a weak  nonlinearity applies to a progressively narrower
range of momenta $k^2 \ll 8\cos^2\theta$ reducing  the range of validity
of the asymptotic expression (\ref{Gkasympt}). Outside that domain one can use
the parabolic form of the magnon dispersion $\epsilon_{\bf k} \approx JSk^2$
to derive another useful asymptotic expression for the decay rate.
In this regime $u_{\bf k}\approx 1$ and $v_{\bf k}=O(\cos\theta)\approx 0$
and  the decay vertex becomes
$\tilde{\Phi}_1({\bf k},{\bf q})  \approx -2$.
The angle between the emitted and the initial magnon can now be large:
$\cos\varphi_0 = q/k$. Performing an analytical integration in Eq.~(\ref{Gk}) and
taking into account that $\cos^2\theta \approx 2(1-H/H_s)$ we obtains
\begin{equation}
\Gamma_{\bf k} \approx 16 \left(1-\frac{H}{H_s}\right)  \ ,
\end{equation}
which is valid for $4\sqrt{1-H/H_s}\ll k \ll 1$.

Modifications to the asymptotic expression (\ref{Gkasympt}) are also expected
in the field regime just above the decay threshold field, $H\rightarrow H^*$.
In this case, $\alpha\to 0$ and the coefficient in
front of $k^3$ in the Born expression for $\Gamma_{\bf k}$ in (\ref{Gkasympt}) diverges.
Such a nonanalytic behavior is nothing but the long-wavelength version of the
decay threshold singularities. The self-consistent regularization (\ref{SCBA})
of Sec.~\ref{selfconsistent} should  remain applicable in this limit.
In order to verify that the power-law behavior (\ref{Gkasympt})
is not modified within the SCBA we substitute a general power-law ansatz for
the decay rate $\Gamma_{\bf k} = \beta k^n$ and assume that
the damping is much larger than the nonlinearity
but still much smaller than the magnon energy:
$\alpha k^3 \ll \beta k^n \ll ck$.  In this case, the decay angle is scaled
as $\varphi^2 \sim k^{n-1}$ with $q\sim k$, which makes the decay vertex
$\tilde{\Phi}_1 ({\bf k},{\bf q})\sim k^{n-3/2}$ instead of (\ref{Phi1simp}).
The power counting on both sides of Eq.~(\ref{SCBA}) yields 
a unique solution: $n=3$.
Therefore, the Born exponent $\Gamma_{\bf k} = \beta k^3$ is not changed
by the self-consistent procedure. The damping coefficient $\beta$
does not diverge near the decay threshold field anymore because
$\alpha$ drops out from the self-consistent equation on $\Gamma_{\bf k}$.
Instead, $\beta$ exhibits a step-like behavior in $(H-H^*)$, which is in 
an accord with the 2D character of the Van Hove singularity at the border of 
the two-particle continuum.

\section{Decays in $XXZ$ SAFM}

\begin{figure}[t]
\includegraphics[width=0.99\columnwidth]{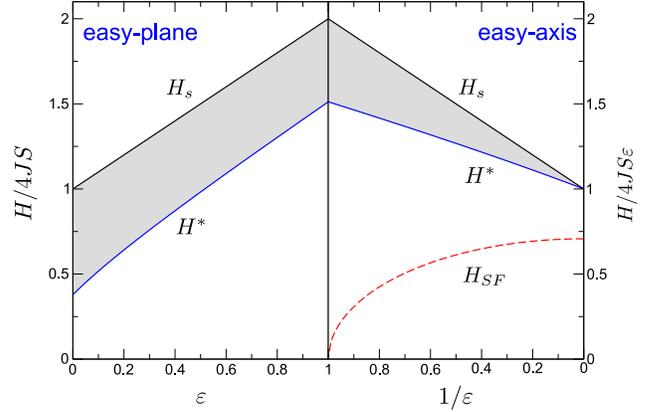}
\caption{
(Color online) Phase diagram of the XXZ model in applied field
as a function of anisotropy parameter $\varepsilon$. The field range where spontaneous magnon decays are allowed is shown in gray.
}
\label{square1}
\end{figure}

The Hamiltonian of the  $XXZ$ antiferromagnet on a square-lattice is
\begin{equation}
\hat{\cal H} = J \sum_{\langle ij\rangle} \left[ S_i^{\perp}S_j^{\perp}
 + \varepsilon S_i^{z}S_j^{z}\right]- H\sum_{i} S_i^{z} \ .
\label{Hxxz}
\end{equation}
The values of the anisotropy parameter $0\leq\varepsilon <1$ describe the
spin system with the easy-plane anisotropy, whereas
$\varepsilon > 1$ corresponds to the easy-axis case.

Minimization of the classical energy yields the transition
field into a fully saturated state
\begin{equation}
H_s = 4JS (1 + \varepsilon) \ .
\label{Hsa}
\end{equation}
For the easy-axis case,
the spin-flop transition between the collinear  state
and  the canted antiferromagnetic state takes place at
\begin{equation}
\label{hsf}
H_{SF}=2JS\sqrt{2(\varepsilon^2-1)}\ .
\end{equation}

The harmonic spectrum in the canted antiferromagnetic state
is given by the same expression  Eq.~(\ref{epsilon})
as for the Heisenberg SAFM with the substitution
$\cos 2\theta \rightarrow (\varepsilon \cos^2\theta - \sin^2\theta)$.
Performing the same type of kinematic analysis as in Sec.~III,
we find that the curvature of the acoustic branch changes at
the threshold field
\begin{eqnarray}
\label{h*}
H^*=4JS\sqrt{(1+\varepsilon)\Bigl(\frac17+\varepsilon\Bigr)} \ .
\end{eqnarray}
Above this field spontaneous two-magnon decays are kinematically allowed.
The phase diagram in the $H$--$\varepsilon$ plane is summarized in
Fig.~\ref{square1}. Note, that in the right panel the units of the field
are different from the left panel.

Several observations are in order:
(i) for arbitrary $\varepsilon$
there is a finite range of fields below the saturation field
where magnons are unstable;
(ii) the spin-flop field $H_{SF}$ in the easy-axis regime is always
below the  decay threshold field $H^*$;
(iii) the easy-plane anisotropy pushes the decay instability to lower
fields compared to the isotropic model.
In the $XY$-limit ($\varepsilon=0$)
the ratio $H^*/H_s=1/\sqrt{7}\approx 0.38$
is two times smaller than for the Heisenberg antiferromagnet.
This might be important for the search of experimental systems where
the phenomenon of magnon decays can be observed.
Note also that the easy-plane anisotropy appears naturally in
the mapping of the BEC-type transition  of the triplet excitations 
in quantum magnets with the singlet ground state onto
the saturation field transition of an effective pseudo-spin-1/2
model.\cite{Mila98} The typical value of the anisotropy parameter
in these problems is $\varepsilon = 0.5$, which implies that 
spontaneous magnon decays must exist in an extended interval of
fields above the condensation field $H_c$.

\end{document}